\documentclass{iau} 
\usepackage{graphicx}

\title[Regulation of SFR] 
{Feedback Regulated Turbulence, Magnetic Fields, and Star Formation Rates in Galactic Disks}

\author[Chang-Goo Kim \& Eve C. Ostriker]   
{Chang-Goo Kim \and Eve C. Ostriker}

\affiliation{Department of Astrophysical Sciences, Princeton University \\
4 Ivy Lane, Princeton, New Jersey, USA 08544 \\ 
email: {\tt cgkim@astro.princeton.edu, eco@astro.princeton.edu}}

\pubyear{2015}
\volume{315}  
\jname{From interstellar clouds to star-forming galaxies: universal processes?}
\editors{A.C. Editor, B.D. Editor \& C.E. Editor, eds.}
\begin{document}

\maketitle

\begin{abstract}
We use three-dimensional magnetohydrodynamic (MHD) simulations to investigate
the quasi-equilibrium states of galactic disks regulated by star formation
feedback. We incorporate effects from massive-star feedback via time-varying
heating rates and supernova (SN) explosions.  We find that the disks in our
simulations rapidly approach a quasi-steady state that satisfies vertical
dynamical equilibrium.  The star formation rate (SFR) surface density
self-adjusts to provide the total momentum flux (pressure) in the vertical
direction that matches the weight of the gas. We quantify feedback efficiency
by measuring feedback yields, $\eta_c\equiv P_c/\Sigma_{\rm SFR}$ (in suitable
units), for each pressure component.  The turbulent and thermal feedback yields
are the same for HD and MHD simulations, $\eta_{\rm th}\sim 1$ and $\eta_{\rm
turb}\sim 4$, consistent with the theoretical expectations. In MHD simulations,
turbulent magnetic fields are rapidly generated by turbulence, and saturate at
a level corresponding to $\eta_{\rm mag,t}\sim 1$. The presence of magnetic
fields enhances the total feedback yield and therefore reduces the SFR, since
the same vertical support can be supplied at a smaller SFR. We suggest further
numerical calibrations and observational tests in terms of the feedback yields.

\end{abstract}

\firstsection 
\section{Introduction}

``What determines the SFR in galaxies?'' In order to answer this long-standing,
fundamental question, a correlation between the SFR and the gas content has
been extensively explored.  Among many studies since the pioneering work by
\cite[Schmidt (1959)]{Sch59}, \cite{Ken98} presents a well-defined power-law
relationship between total gas surface density ($\Sigma$) and the SFR surface
density  ($\Sigma_{\rm SFR}$) for galaxies as a whole, $\Sigma_{\rm SFR}\propto
\Sigma^{1+p}$ with $p=0.4$. This observed correlation was soon widely accepted
as the ``Kennicutt-Schmidt law'' (KS law) and used as a star formation recipe for
large scale galaxy formation and cosmological simulations.

The observed power-law index with $p=0.4$ of the KS law makes it tempting to
infer simple dimensional relationship, $\Sigma_{\rm SFR}=\Sigma/t_{\rm dep}$,
with the gas depletion time related to the gas free-fall time, $t_{\rm
dep}\propto t_{\rm ff}\sim(G\rho)^{-1/2}$.  With a fixed gas scale height, this
relation would imply $p=0.5$, close to the observed value. Many theoretical
studies based on this simple argument have been investigated, with low star
formation efficiency per free-fall time $\epsilon_{\rm ff}\equiv t_{\rm
ff}/t_{\rm dep}\sim 1\%$ (e.g., \cite{Kru12}).  On scales of molecular clouds,
the low efficiency has been attributed to the broad density probability
distribution function generated by supersonic turbulence (e.g., \cite{Pad14}
and references therein), but it is unclear whether this picture can be simply
extended to large scales ($>$0.1-1 kpc). 

Moreover, recent high-resolution observations of nearby galaxies reveal more
complex correlations (\cite{Big08,Ler08}). In particular, the simple power-law
relation between $\Sigma_{\rm SFR}$ and $\Sigma$ fails at low surface density
regime ($\Sigma < 10 \; M_{\odot}\; {\rm pc}^{-2}$), where the gas is
predominantly atomic. Rather, the power-law index becomes steeper and/or varies
from one galaxy to another. The SFR shows a tighter correlation with the
stellar surface density $\Sigma_*$ or its combination with $\Sigma$ (e.g.,
\cite{Ler08}).

The increasing complexities of the observed KS law at low-$\Sigma$ regime
implies that $\Sigma$ is not the only control parameter of the star formation.
In this article, we describe a fundamental correlation based on physical
causality between the SFR surface density and the total pressure ($P_{\rm
tot}$). We in \S~\ref{sec:theory} and \S~\ref{sec:simulation} respectively
summarise the theory from \cite{Ost10,Ost11,Kim11} and simulations from
\cite{Kim11, Kim13, Kim15}.

\begin{figure}
\begin{center}
 \includegraphics[width=0.7\textwidth]{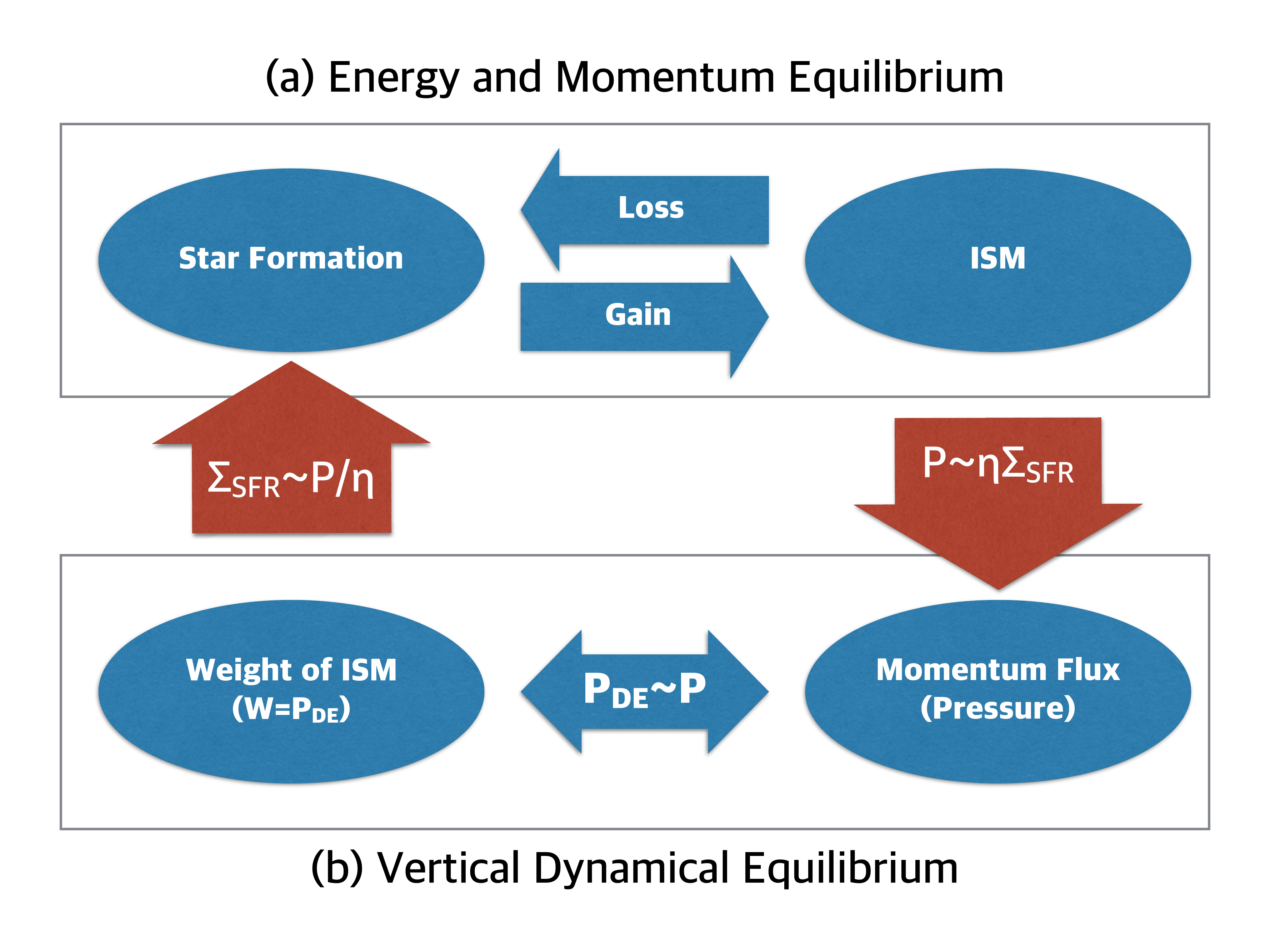} 
 \caption{Schematic diagram of the equilibrium theory. (a) Energy/momentum
equilibrium sets the total pressure in response to the SFR, $P_{\rm
tot}\sim\eta\Sigma_{\rm SFR}$. (b) Vertical dynamical equilibrium constrains
the total momentum flux (pressure), $P_{\rm tot}\sim P_{\rm DE}$, and hence the
SFR, $\Sigma_{\rm SFR}\sim P_{\rm DE}/\eta$.}
\label{fig1}
\end{center}
\end{figure}

\section{Theory} \label{sec:theory}

The interstellar medium (ISM) disk in an equilibrium state should satisfy
vertical force balance between gravity and pressure gradients, which can be
directly derived from the momentum equation of MHD.\footnote{ Although the
resulting equation would be essentially the same with so called
\textit{hydrostatic equilibrium} with an effective (total) sound speed, we
prefer to term this \textit{vertical dynamical equilibrium} since the ISM is
highly dynamic and equilibrium holds only in an average sense.}  In the
integrated form, the vertical dynamical equilibrium can be written as
$\mathcal{W}=\Delta P_{\rm tot}$, a balance between the total weight of gas and
the momentum flux differences across the gas disk.

The other condition to satisfy is the energy/momentum equilibrium between gain
from star formation feedback and loss in the dissipative ISM.  Since cooling
and turbulence dissipation time scales are typically short compared to
dynamical time scales, continuous injection of energy and momentum is
necessary to heat gas and drive turbulence. The far-UV radiation from massive
young stars is the major heating source in the atomic ISM via the photoelectric
effect onto grains. The momentum injection from SNe is the dominant source of the
turbulence driving. Because the energy and momentum from stellar radiation
and SNe fundamentally derive from nuclear processes, SF feedback is a highly
efficient way to balance losses in the ISM. 

Figure~\ref{fig1} shows a schematic diagram for (a) energy/momentum equilibrium
and (b) vertical dynamical equilibrium as well as the connection between two
equilibria.  Since the ISM mostly gains energy/momentum through the star
formation feedback, the total pressure (momentum flux) is set by the SFR
surface density. The total pressure determined by Figure~\ref{fig1}(a) provides
the vertical support in Figure~\ref{fig1}(b), which should match with the
dynamical equilibrium pressure $P_{\rm DE}$ (or the weight of the gas
$\mathcal{W}$).

The equilibrium is naturally a stable one.  For example, if the SFR gets higher
than equilibrium, enhanced thermal heating and turbulence driving set the total
pressure higher than the equilibrium level. The disk becomes thermally and
dynamically hotter, vertically expanding and dispersing cold, dense clumps to
quench further star formation. On the other hand, when the SFR drops below of
the level of equilibrium, reduced feedback makes the disk thermally and
dynamically cold and susceptible to gravitational collapse, forming more stars.
Therefore, the SFR is self-regulated to satisfy both equilibria shown in
Figure~\ref{fig1}.

In order to quantify this process, we define the \textit{feedback yield} of any
pressure component ``$c$'' (= thermal, turbulent, or magnetic) in
suitable units as
\begin{equation}
\eta_c \equiv \frac{P_{c,3}}{\Sigma_{\rm SFR,-3}},
\end{equation}
where $P_{c,3}\equiv P_{c}/(10^3 k_B {\rm \; cm}^{-3}{\rm \; K})$ and
$\Sigma_{\rm SFR,-3}\equiv \Sigma_{\rm SFR}/(10^{-3} \; M_{\odot}{\rm \;
pc}^{-2} {\rm \; Myr}^{-1})$.  The feedback yield can be considered as a
energy/momentum conversion efficiency of the star formation feedback, depending
on the detailed thermal and dynamical processes in the ISM. To first order, we
can simply connect the thermal and turbulent pressures with the SFR surface
density linearly.  Our adopted cooling and heating formalism gives $\eta_{\rm
th}=1.2$, and the momentum feedback prescription with specific momentum
injection per star formation $(p_*/m_*)=3000 {\rm \;km/s}$ gives $\eta_{\rm
turb}=3.6$.

\section{Numerical Simulations}\label{sec:simulation}

Utilizing the \textit{Athena} code (\cite{Sto08}), we run three-dimensional
simulations for the local patch of galactic disks including optically thin
cooling, galactic differential rotation, self-gravity, vertical external
gravity, and magnetic fields. We apply a spatially-constant, time-varying
heating rate $\Gamma\propto \Sigma_{\rm SFR}$, and also momentum injection from
SNe $\propto \Sigma_{\rm SFR}$.  While in some other recent simulations, the
SFR and SN rate is pre-specified and constant in time, for all of our models
the time-dependent SFR and SN rate are self-consistently set by
self-gravitating localized collapse.

Our simulations achieve a quasi-steady state after a few vertical oscillation
times (less than one orbit time). We confirm vertical dynamical equilibrium
using horizontally and temporally averaged vertical profiles that are converged
for different initial and boundary conditions as well as numerical resolutions.
For a wide range of disk conditions such that $0.1<\Sigma_{\rm SFR,-3}<10$, the
equilibrium thermal and turbulent pressures give consistent feedback yields,
$\eta_{\rm th}=1.3\Sigma_{\rm SFR,-3}^{0.14}$ and $\eta_{\rm turb}=
4.3\Sigma_{\rm SFR,-3}^{0.11}$, respectively.

In MHD simulations, the time scales to reach a quasi-steady state depend on the
initial magnetizations. For initial magnetic energy varying by two orders of
magnitude, however, saturated states converge to the same asymptote for the
turbulent magnetic fields, whose energy is about a half of the turbulent
kinetic energy. The final turbulent magnetic fields provide additional vertical
support that is directly related to the turbulent pressure and hence the SFR,
giving rise to $\eta_{\rm mag,t}\sim1$ for solar neighborhood models. Since we
fix disk parameters for MHD models, further investigation is necessary to
calibrate the detailed dependence of $\eta_{\rm mag,t}$ on $\Sigma_{\rm SFR}$.

\section{Concluding Remarks}

We have developed a theory for self-regulation of the SFR based on
energy/momentum equilibrium and vertical dynamical equilibrium, and confirmed
and calibrated this theory with numerical simulations. The former equilibrium
gives rise to the correlation between $\Sigma_{\rm SFR}$ and $P_{\rm tot}$
owing to the physical causality; the higher/lower SFR causes higher/lower
energy density and momentum flux ($P_{\rm tot}=\eta\Sigma_{\rm SFR}$).  The
latter equilibrium sets $\Sigma_{\rm SFR}$ based on the requirement $P_{\rm
tot}=P_{\rm DE}$.  Therefore, correlations between $\Sigma_{\rm SFR}$ and
galactic properties are caused by dependences embedded in $P_{\rm DE}$ such
that
\begin{equation}\label{eq:PDE}
P_{\rm DE}\equiv \mathcal{W} \approx \frac{\pi G \Sigma^2}{2}+
\Sigma\sigma_z(2G\rho_{\rm sd})^{1/2}
\end{equation}
where $\rho_{\rm sd}$ is the midplane density of stars plus dark matter and
$\sigma_z$ is total vertical velocity dispersion. The observed complexities in
the KS law for the low-$\Sigma$ regime naturally arise when the second term in
RHS of Equation (\ref{eq:PDE}) dominates.

Lastly, we suggest further numerical calibrations and observational tests of
the equilibrium theory. Theorists who include any form of star formation
feedback can calibrate $\eta_c$ from a ``$P_c$-$\Sigma_{\rm SFR}$'' plot for
each measured pressure (turbulent, thermal, magnetic) in their simulations. It
could be interesting tu check consistency and/or to investigate differences
among simulations with different setups, including comparing global {\it vs.}
local models. Additional calibrations of other components such as radiation
pressure and cosmic ray pressure would enable comparison of the relative
importance of each component.

It is difficult to measure the total pressure directly from observations even
in solar neighborhood. However, the dynamical equilibrium pressure can be
determined from direct observables (such as those from \cite{Ler08}) with
proper assumptions.\footnote{Please refer to the following link for
a practical guide: {\tt
http://www.astro.princeton.edu/$\sim$cgkim/to\_observers.html}} From the
``$P_{\rm DE}$-$\Sigma_{\rm SFR}$'' observed plot, one can measure total
feedback yield $\eta=P_{\rm DE,3}/\Sigma_{\rm SFR,-3}$.  This can be compared
with the sum of the theoretical values $\eta_c$ as a test of the
self-regulation theory, also constraining the dominant sources of SF feedback.

\end{document}